\documentclass[prl,aps,twocolumn,superscriptaddress]{revtex4}
\usepackage[dvips]{graphicx}
\usepackage{txfonts}

\begin{document}
\title{Luminosity, selfgravitation and nonuniqueness of stationary   accretion}
\author{Janusz Karkowski}
\author{Edward Malec}
\author{Krzysztof Roszkowski}
\affiliation{M. Smoluchowski Institute of Physics, Jagiellonian University, Reymonta 4, 30-059 Krak\'{o}w, Poland}

\begin{abstract}
We investigate newtonian description of accreting compact bodies with hard surfaces,
including luminosity  and selfgravitation of  polytropic perfect fluids.  
This nonlinear integro-differential problem reduces, under appropriate boundary 
conditions, to an algebraic relation between luminosity and the 
gas abundance in  stationary spherically symmetric flows. There exist, for a given luminosity,
asymptotic  mass and the asymptotic temperature,  two sub-critical solutions 
that bifurcate from a critical point. They  differ by the fluid content and 
 the mass of the compact centre. 

\end{abstract}

\maketitle
 
\section{Introduction.}
 
We study in this paper a   newtonian accreting systems consisting of a compact body
enclosed by a ball of selfgravitating gas. Our aim   is to investigate 
the following {\it inverse problem}: assume that one knows the luminosity, total mass,
asymptotic temperature (or, equivalently, the asymptotic speed of sound)
  and   the equation of state of the gas.
Assume that the gravitational potential has a fixed value at the boundary of a compact body, but the 
radius of the body is arbitrary to some extent. The question arises: can one   
specify the mass of the  compact body?  And if one can, is the specification unique? 
That issue is interesting  in itself within the framework  of astrophysics but it is also
related to  following  fundamental physical aspect. 

The standard point of view that  compact objects of a mass exceeding several 
solar masses  constitute black holes has been challenged by Mazur and Mottola \cite{Mazur}, who developed 
the notion of gravatars --- stars almost as compact as black holes but dispossessed 
of event horizons. Gravastars are supposed  to be  built from matter which  violates energy 
conditions \cite{Hawking} and therefore they can avoid the standard compactness limits (like the Buchdahl 
theorem \cite{Buchdahl}) or mass  bounds. The concept of gravastars originated from the
belief that  the only  place after Big Bang that   quantum gravity could manifest is the 
collapse of matter  to a pre-(event horizon) phase, and that quantum gravity effect may or can 
actually stop the collapse. Yet another conclusion, if one believes in this argument, is that
the search for quantum gravity should start from the revision of the scenario of 
the gravitational collapse to a black hole. 

The issue of the validity of gravastars to astrophysics can be verified, at least in principle,
by observations. The fundamental question is whether  observations can distinguish 
compact bodies endowed with a hard surface (gravastars) from black holes. 
Narayan \cite{Narayan}  presented observational arguments in favour of the
existence of black holes. Abramowicz, Klu\'zniak and Lasota \cite{Abramowicz} 
raised several objections, expressing their view that present accretion disk models are not 
capable to make distinction between accretion onto a black hole or a gravastar.

We choose below  the  simplest selfcontained model  that can be
interpreted as a radiating system. While we do not address  directly the problem 
of distinguishing between a gravastar and a black hole, our results show that there 
is an ambiguity even in the newtonian level.   There can exists at least two  
solutions (two compact stars with a hard surface) to the {\it inverse problem}
 described above.  This is not quite unexpected. In a recent general-relativistic
analysis (valid also in the newtonian limit) the mass 
accretion rate  $\dot M$ behaves  like $y^2(1-y)$, where  $y=M_\ast /M$ 
is the ratio of (roughly) the mass of the compact core to the total mass \cite{Kinasiewicz}.
Thus  there  exists a maximum of $\dot M$ when $y=2/3$.  The luminosity equals to the
mass accretion rate $\dot M$ multiplied by the available energy per unit mass
--- the potential $\phi (R_0)$
at the hard surface of a compact star.  Hence this simple
analysis of accretion would suggest the  existence of two  weakly luminous    regimes:
one   rich in fluid  with $m_f\approx M $  and the other with a small amount
of fluid, $m_f/M\ll 1$. To a given luminosity $x$ might have corresponded two systems
with different $y$'s.  The actual situation in the case with the luminosity is less clear,
because the luminosity impacts also the accretion rate $\dot M$ and there exists
a complex algebraic relation $y=y(x)$, but the nouniqueness is still retained.

\section{The Shakura model.} 

Stationary accretion of spherically symmetric fluids with
luminosities close to the Eddington limit has been investigated since
the pioneering work of Shakura \cite{Shakura}. In \cite{Shakura} the gas pressure
and its selfgravity have been ignored and the analysis has been purely newtonian.
Later researchers  included both these aspects \cite{later} and extended the analysis 
onto relativistic systems (\cite{Thorne} --- \cite{Rezzola}). 
 We will consider a spherically symmetric compact ball of a fluid falling onto a non-rotating 
compact body.  
The areal velocity $U$ of a comoving particle labelled by coordinates $(r,t)$
 is given by $U(r,t) = \partial_t R $, where $t$  is a comoving (Lagrangian) time.
 $p$ denotes pressure, $L(R)$ and $L_E$ are the local and 
Eddington luminosities, respectively.
 The quasilocal mass
$m(R)$ is defined by $\partial_R m(R) = 4 \pi R^2 \varrho $. The mass
accretion rate is  
\begin{equation}
\dot M = -4 \pi R^2 \varrho U.
\label{dot}
\end{equation}
  $\varrho $ is the baryonic mass density.
The equation of state is taken to be  polytropic,
   $p = K \varrho^\Gamma$ with $\Gamma$ being a constant, ($1 < \Gamma \le 5/3$).
We assume  a steady collapse of the  fluid, which means that all its 
characteristics are constant at a fixed areal radius $R$: 
$\partial_t X|_{R = \mathrm{const}} = (\partial_t - (\partial_t R) \partial_R) X = 0$,
 where $X = \varrho, U, a^2$. $a=\sqrt{\partial_\rho p} $ is the 
speed of sound. Strictly saying, a stationary accretion must lead  
to the increase of the central mass. This  
in turn means that the notion of the "steady accretion" is approximate ---
it demands that the mass accretion rate is small and  the time scale is  short,
so  that the quasilocal  mass  $m(R)$  does not change significantly.
The total mass $m(R_\infty )$ will be denoted by $M$.
Furthermore, we assume that the radius $R_\infty$ of the ball of fluid and other
boundary data are such that $U^2_{\infty} \ll \frac{m(R_\infty)}{R_\infty}
 \ll a^2_\infty$.  

The  steadily accreting gas is described by a system of
 integro-(ordinary)-differential equations.
They consist of the (Euler) momentum conservation equation
\begin{equation}
U\partial_R U = -{Gm(R)\over R^2}  -{1\over \varrho } \partial_Rp +  
  \alpha  {L(r) \over R^2}, 
\label{1}
\end{equation}
the mass conservation 
\begin{equation}
\partial_R \dot M =0
\label{2}
\end{equation}
 and the energy conservation
\begin{equation}
L_0-L(r) = \dot M \left( {a^2_\infty \over \Gamma -1} - {a^2 \over \Gamma -1}
-{U^2\over 2} -\phi (R) \right) .
\label{3}
\end{equation}
The constant $\alpha $ is defined by $\alpha =\sigma /\left( 4\pi m_p c\right) $,
$L_0$ is the total luminosity  and $\phi (R) $ is the newtonian gravitational potential.  
Strictly saying, in Eq. (\ref{3}) should appear 
${a^2_{\infty } \over \Gamma -1}+{U^2_\infty \over 2}- {M\over R_\infty }$
 but this is well approximated by ${a^2_\infty \over \Gamma -1}$,
due to the aforementioned boundary conditions.  
 
We define the Eddington luminosity as that at the outermost layer of the accreting 
gas, that is as  $L_E=GM/\alpha $, while the total luminosity $L_0$
is equal to the product of the mass accretion rate and the total available energy
per unit mass $\phi_0\equiv |\phi (R_0)|$, where $R_0$ is an areal 
radius of the boundary of the compact body,
$L_0=\dot M \phi_0$ (\cite{Shakura}). The quantity $\phi_0$ is   fixed  
and $R_0$ in the constructed configurations is by definition
the areal radius at which the absolute value of the surface potential equals  to  $\phi_0$. 
Differentiating $\dot M$ with respect to $R$, using the equality 
$\partial_Ra^2/(\Gamma -1) = a^2\partial_R\ln \varrho =\partial_Rp/\varrho $ and 
combining the obtained equation with (\ref{1}) yields the differential equation
$
\partial_R\ln L = {\alpha \dot M \over R^2}.
$
Its solution can be written in terms of the Eddington and total luminosities as follows

\begin{equation}
L = L_0\exp \left( {-L_0\tilde R_0\over L_ER}\right) .
\label{5}
\end{equation}
Here $\tilde R_0 \equiv GM/|\phi (R_0)|$ is a kind of modified size measure  of the compact  
body. In the case of test fluids $\tilde R_0=R_0$. In the case of gravastars one should 
assume that $|\phi (R_0)|\sim c^2/2$ and $\tilde R_0 \sim 2GM/c^2$ is the Schwarzschild radius;
 while gravastars 
would require the general-relativistic treatment, we believe that essential features 
will be the same as in the newtonian description. 
 
\section{The mass accretion rate.}
 
In what follows it is assumed that there exists a sonic point, but  we believe that the same 
(nonuniqeness) result would have been obtained for subsonic solutions.    
The sonic horizon (sonic point) is   a location where
   $|U| = a$.  In the following we will denote by the asterisk all values referring 
to the sonic points, e.g. $a_\ast$, $U_\ast $, etc. Differentiation 
with respect the areal radius will be denoted as prime $'$. One finds from the mass conservation
$U'=-U\left( \left(\varrho '/\varrho \right) +2/R\right) $. Inserting  that into the 
Euler equation (\ref{1}), one finds
\begin{equation}
{\varrho '\over \varrho } (a^2-U^2)= {2\over R} \left(
U^2-{Gm\left( R\right) \over 2R }+{L\left( r\right) \alpha \over 2R} \right) .
\label{6}
\end{equation}
thus at the sonic point the three characteristics, $a_\ast$, $U_\ast$ and $M_\ast / R_\ast$
are related, 
\begin{eqnarray}
a_\ast^2  &=& U_\ast^2 ={Gm\ast \over 2R_\ast }\left( 1- {L_\ast \alpha \over GM_\ast }\right) =
\nonumber\\
&&{Gm\ast \over 2R_\ast }\left( 1- {L_\ast  M \over L_EM_\ast }\right).
\label{7}
\end{eqnarray}
It is clear from (\ref{7}) that in order to have a critical flow one should have 
\begin{equation}
{L_\ast  M\over  L_EM_\ast }<1.
\label{cf}
\end{equation}
It is useful to express the velocity $U$ and the mass density $\varrho $ in the following way
\begin{equation}
U=U_\ast \frac{R^2_\ast}{R^2}
\left(  {a^2_\ast \over a^2}  \right)^{1/(\Gamma -1)},
\label{U}
\end{equation}
where $U_\ast$ is the negative square root, and 
\begin{equation}
\varrho = \varrho_{\infty } \left( a/a_\infty \right)^{2/(\Gamma - 1)},
\label{rho}
\end{equation}
where the constant  $\varrho_\infty$ is equal to the mass density of  collapsing
fluid at the boundary $R_\infty $. There exist four stationary branches crossing
at the sonic point, one of which is the accretion flow. 

Define  
\begin{eqnarray}
&&x \equiv {L_0\over L_E},~~~~y\equiv {M_\ast \over M }, 
~~~~  \gamma \equiv {\tilde R_0\over R_\ast }, \nonumber\\
&& \Delta_*\equiv L_0-L_\ast -2\dot M a^2_\ast {L_\ast \over yL_E- L_\ast },\nonumber\\
&& \Psi_\ast \equiv -\phi_\ast -{GM_\ast \over R_\ast}.
\label{def}
\end{eqnarray}
In new variables the necessary condition (\ref{cf}) for the existence of a sonic 
point reads $x\exp \left( -x\gamma  \right) <y$.
A straightforward algebra leads to the following equation
\begin{equation}
a^2_\infty -{5 - 3 \Gamma \over 2}a^2_\ast =\left( \Gamma -1\right) 
\left( {\Delta_\ast \over \dot M} -\Psi_\ast \right) .
\label{theorem2a}
\end{equation}
We shall assume that $\gamma <1 $ and $x\gamma \ll 1$.
From Eq. (\ref{7}) follows now   
\begin{equation}
U^2_\ast ={GM_\ast \over 2R_\ast }\left( 1- { x\over y }\exp \left( -x\gamma  \right)\right) .
\label{y}
\end{equation}
Then $\Delta_\ast =0$ up to the first order while 
$\Psi_\ast =4G\pi \int_{R_\ast }dr r \varrho $. One can show, using arguments 
similar to those
applied later, that $|\Psi_\ast | \ll a^2_\ast$.
Therefore (\ref{theorem2a}) simplifies to
\begin{equation}
{a^2_\ast \over a^2_\infty } =  {2\over 5 - 3 \Gamma };
\label{theorem2}
\end{equation}
that is exactly the same result as in the Bondi model \cite{Bondi}.
A careful reader would notice that the above  derivation requires
 that $\Gamma $ is somewhat  smaller than 5/3.

The  rate of  the  mass accretion $\dot M$ as  given by (\ref{dot}) 
can be expressed in terms of the characteristics of the sonic point. One obtains,
using equations (\ref{7}) and (\ref{dot}), that  

\begin{equation}
\dot M = G^2\pi^2 M^2{\varrho_\infty \over 
a^3_\infty} \left( y-x\exp \left( -x\gamma  \right) \right)^2\left( \frac{a_\ast^2}{a_\infty^2} 
\right)^\frac{(5 - 3 \Gamma)}{2(\Gamma - 1)} .
\label{dotm}
\end{equation}
Now, one can can show in a way similar to that applied in the case of relativistic 
accretion (\cite{Kinasiewicz}) that under the previously assumed 
conditions and $R_\ast >> \tilde R_0$ one has $\varrho_\infty =
\chi_\infty \left( M-M_\ast \right) =M\chi_\infty \left( 1-y\right) $ for $\Gamma \in (1, 5/3 -\delta )$
with some small $\delta $.
 
More specifically, one needs to show that the energy conservation equation (\ref{3})
yields
\begin{equation}
{a^2(R)\over a^2_\infty } < 1-\left( \Gamma -1 \right) {\phi (R)\over a^2_\infty };
\label{a2}
\end{equation}
bounding $|\phi (R) |$ from above by $GM/R$
one obtains  
\begin{equation}
{a^2(R)\over a^2_\infty } < 1+\left( \Gamma -1 \right) {GM \over Ra^2_\infty }.
\label{a3}
\end{equation}
Therefore  (\ref{rho}) implies
 $\varrho \le \varrho_\infty \left( 1+ {GM \over Ra^2_\infty }\right)^{1/(\Gamma - 1)} $.
The conclusion follows now from using this bound in $M-M_\ast =\int_VdV \varrho $ 
and appealing to the asymptotic conditions laid down in the forthcoming text. 
The proportionality constant
$\chi_\infty $ is roughly the inverse of the volume of the gas outside of the sonic sphere.
  Since the total luminosity $L_0=\phi (R_0) \dot M$ and $a^2_\ast / a^2_\infty $ 
is given by (\ref{theorem2}), one obtains  that 

\begin{equation}
L_0=\phi_0  G^2\pi^2 {\chi_\infty  M^3 \over 
a^3_\infty}\left( 1-y\right) \left( y-x\exp \left( -x\gamma 
 \right) \right)^2\left(   {2\over 5 - 3 \Gamma }
\right)^\frac{(5 - 3 \Gamma)}{2(\Gamma - 1)}.
\label{L1}
\end{equation}

\section{The analysis of the luminosity equation.}

It is convenient to cast the formula (\ref{L1})
 in terms of the relative luminosity, $x=L_0/L_E$,
\begin{equation}
x= \beta \left( 1-y\right) \left( y-x\exp \left( -x\gamma  \right) \right)^2,
\label{l1}
\end{equation}
where $\beta = \chi_\infty \phi_0 \alpha  G\pi^2  {M^2 \over  a^3_\infty} \left( {2\over 5 - 3 \Gamma }
\right)^\frac{(5 - 3 \Gamma)}{2(\Gamma - 1)}$.

We shall analyze solutions of this cubic (with respect to the variable $y$)
equation. One can show  that there exist at least two solutions $y(x,\beta )$ for any parameter $\beta $, 
 $0\le \gamma <1$ and  the relative luminosity $x$ smaller than a certain critical 
value $a$. We shall add the adjective "critical"
 to all characteristics of this critical point; hence a critical luminosity 
is that corresponding to $x=a$. The  obtained results are following.

{\bf Theorem.} Define $F(x,y)\equiv x-  
\beta \left( 1-y\right) \left( y-x\exp \left( -x\gamma  \right) \right)^2$. Then

i)  there exists
a critical point   $x=a, y=b$ of $F$, that is $F(a,b)=0$ and $\partial_yF(x,y)|_{a,b}=0$, 
 with $a$ and $b$ satisfying  $0<a <b<1$ 
and   $b={2+a\exp \left( -a\gamma  \right) \over 3}, a=4\beta (1-b)^3$;

ii) For any $0<x<a$ there exist two solutions $y(x){^+_-}$ bifurcating from $(a,b)$.
They are locally approximated by  formulae
\begin{equation}
y{^+_-}=b\pm {\sqrt{(a-x)(b+a\exp (-a\gamma )(1-2a\gamma ))}
\over \sqrt{\beta (b-a\exp (-a\gamma )) (1-a\exp (-a\gamma ))}}.
\label{bif}
\end{equation}

iii) The relative luminosity $x$ is extremized at the critical point $(a,b)$.

{\bf Proof.}  

The two criticality conditions yield two equations
\begin{eqnarray}
&&a-\beta (1-b) (b-a\exp (-a\gamma ))^2=0, \nonumber\\
&& b-a\exp (-a\gamma )=2(1-b).
\label{e}
\end{eqnarray}
From Eqs (\ref{e}) one immediately obtains the desired expressions
\begin{equation}
b={2+a\exp 
\left( -a\gamma  \right) \over 3},~~~~
a=4\beta (1-b)^3.
\label{e00}
\end{equation}
Inserting the second of eqs (\ref{e00}) into the second equation in (\ref{e}), one arrives at
\begin{equation}
b={2\over 3} + {4\beta \over 3}(1-b)^3\exp (-4\beta (1-b)^3\gamma ) .
\label{e1}
\end{equation}
Both sides of this equation are continuous functions of $b$ and at $b=0$ the right
hand side of (\ref{e1}) is bigger than the left hand side, while at $b=1$ the opposite
holds true. Therefore there exists a solution. A closer inspection shows that there is
only one  critical solution. This fact actually guarantees that  solutions $y(x)$ that
bifurcate from $(a,b)$ extend onto  the whole interval $x\in (0, a)$. (Hint: use the 
implicit function theorem and the fact that $(a,b)$ is the unique critical point).  
The specific form of a solution close to a critical point can be obtained as 
follows. Insert $x=a+\epsilon , y=b + y(\epsilon )$ into $F(x,y)=0$ and expand $F$ 
keeping the terms of lowest order. One obtains a reduced algebraic equation
\begin{eqnarray}
&&\left( 1+2\beta \left( 1-b \right) \left( b-ae^{-a\gamma }\right)
\left( ae^{-a\gamma }\gamma -e^{-a\gamma } \right) \right) \epsilon +\nonumber \\
&&
\beta \left( 3b- 1- 2 ae^{-a\gamma } \right) y^2(\epsilon )=0.
\label{Lya}
\end{eqnarray}
Notice that at the critical point $3b- 1- 2 a\exp (-a\gamma )
=1- a\exp (-a\gamma ) $ (see the second of Eqs (\ref{e}) ) while 
\begin{eqnarray}
&&
2\beta \left( 1-b \right) \left( b-a\exp (-a\gamma )\right)
\left( a\exp (-a\gamma )\gamma -\exp (-a\gamma ) \right) = \nonumber\\
&&
2a\left( a\exp (-a\gamma )\gamma -\exp (-a\gamma ) \right)/(b-a\exp (-a\gamma ).
\label{e2}
\end{eqnarray}
Inserting (\ref{e2}) into (\ref{Lya}) and finding $y(\epsilon )$ from the latter
immediately leads to the approximate solution of ii).   

Now we shall  prove the third part of the Theorem. Let $(x_0,y_0)$ be a non-critical 
solution of $F(x,y)=0$ with the domain being
subset $xe^{-x}<y$ of the square $0<x<1, 0<y<1$. Then 
from the implicit function theorem there exists a curve $x(y)$ such that 
$F(x(y),y)=0$ for $y$ belonging to some vicinity of $y_0$. Along this curve one has
\begin{equation}
{dx\over dy}=\beta {\left( y-xe^{-x\gamma }\right) \left( 3y-2- xe^{-x\gamma }
\right) \over
1+ 2\beta (1-y) (y-xe^{-x\gamma }) (e^{-x\gamma }-\gamma xe^{-x\gamma }) }.
\end{equation}
 The denominator 
is strictly positive while the nominator  vanishes only at critical points. 
That proves the assertion.  It follows from the form of the approximated solution constructed in 
part ii)   that $x=a$ is the extremal value of the relative luminosity.
That ends the proof of the Theorem.

It is easy  to observe that the coordinates $a,b$ of the critical point increase
with the increase of  the parameter $\beta $. Indeed, from Eq. (\ref{e00}) follows
${da\over db}=e^{-a\gamma }(1-a\gamma )$; this is bigger than zero, since $a\gamma <1$.
Differentiation of the first equation in (\ref{e00}) with respect to $\beta $ yields 
${db\over d\beta }=4(1-b)^3/({da\over db} +12 \beta (1-b)^2)>0$; we exploited
here the fact that ${da\over db}>0$. Since along the critical curve ${da\over d\beta }
=  {da\over db }{db\over d\beta }$, we obtain also ${da\over d\beta }>0$.
In the particular case of   $\gamma =0$ one can explicitly find $b$ by solving (\ref{e1}),
\begin{equation}
b=1- {\left( \beta^2+\beta^{3/2} \sqrt{1+\beta  }\right)^{2/3}-
\beta \over 2\beta \left( \beta^2+\beta^{3/2} \sqrt{1+\beta  }\right) ^{1/3}};
\label{cubic}
\end{equation}
(\ref{e00}) gives then $a(\beta )$ and one can check explicitly that both $a$ and $b$ 
monotonically increase with the increase of $\beta $.
The forthcoming figure shows the two branches $x_1(y), x_2(y)$ bifurcating from 
a critical point, in the  case when $\gamma =0$.
  
The next interesting fact is that at  critical points the parameter $b$ is not smaller
than 2/3. This lower bound is saturated at  small $\beta $, i. e., 
when  the relative luminosity $a$ is small (notice that $a<\beta $), which   corresponds to the
maximal rate of the mass accretion in irradiating systems \cite{Kinasiewicz}.
Since the  fluid abundance is equal to $1-b$, we can conclude that critical configurations
have less fluid than 1/3 of the total mass, and the upper bound 1/3 is saturated in the
limit of vanishing radiation. The maximal mass accretion rate does not occur exactly at
$y=2/3$, as  in systems with no radiation  \cite{Kinasiewicz}, but at 
a  somewhat larger value.    

\begin{figure}
\includegraphics[height=7cm]{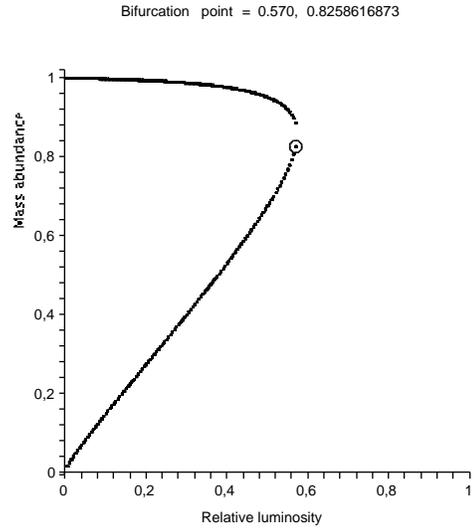}
\caption{\label{figure1}Here $\beta =50$. 
The mass abundance is put on the ordinate while
the relative luminosity $x$ is shown on the abscissa. The circle encloses
the bifurcation point $(a,b)$.}
\end{figure}
  
\section{Discussion.}

In the foregoing text we have assumed the existence of an accreting system
satisfying particular boundary conditions. Although the detailed behaviour 
is given by the integro-differential nonlinear system (\ref{1}-\ref{3}),
its essential features can be deduced from  the algebraic equation (\ref{l1}).
The validity of this  projection of the original equations   to the algebraic
problem relies on the existence of appropriate solutions. We checked numerically
that there do exist such solutions and that the relative error made in the above 
approximations is of the order of $10^{-3}$. Here follows an example.

The mass of the system is 100 solar masses ($M=1.989\times 10^{35}$ grams),
its size $1.254\times 10^{14}$ cm and the relative luminosity $x\equiv {L_0\over L_E}= 0.1$.
The adiabatic index was taken $\Gamma = 4/3$, the asymptotic speed  of sound
$a^2_\infty = 1.875\times 10^{7} $ cm/s and $\tilde R_0=6\times 10^7$ cm.
Let us remark that if  $M$ and $\tilde R_0$ are fixed, then also the potential at the 
surface of the compact core is fixed. The set of quantities hitherto specified 
should in principle determine a solution; any additional information 
would lead to a contradiction.

We have found, by integrating the original integro-differential equations
(\ref{1} --- \ref{3}) that there exist two solutions with following characteristics:
i) solution  I --- $R_\ast =1.2\times 10^5$ cm, $M_\ast =0.11 M$ and $\varrho_\infty
 \approx 2.144 \times 10^{-8}$ g/cm$^3$;
ii)  solution  II --- $R_\ast =1.068 \times 10^7$ cm, $M_\ast =0.9999 M$ and $\varrho_\infty
 \approx 2.41 \times 10^{-12}$ g/cm$^3$.
The  only difference between numerical solutions and algebraic approximations shows
in $\varrho_\infty$ and it is of the order of 0.1\% .
  
Under a simplifying assumption that $\gamma \ll 1 $ one can 
 neglect the exponential term in  $x\exp (-x\gamma )$. The analysis of (\ref{e00}) and
(\ref{cubic}) allows us to conclude that $a$ can be as close as one wishes to 1
for large $\beta $. Putting that in physical terms, the luminosity $L_0$ can be close 
to the Eddington luminosity $L_E$ if the parametr $\beta $ is large enough, or more precisely,
if the product $(M/M_0)^2/(a_\infty /a_{0\infty })^3 $ (where $M_0$ and $a_{0\infty }$ are
some reference quantities) is large enough. The two bifurcation solutions have a luminosity
$x<a$ and they are characterized by two different numbers $y_1, y_2$; the latter can be markedly different
(say  $y_1>10y_2$) only for $x\ll a$. This  is suggested  by  the point ii) of the
Theorem   which shows the  branching of the two solutions
from a critical point, but numerical calculations confirm that more precisely.
A practical conclusion is a follows:  if a given system radiates with
the luminosity close to its critical luminosity (and, in particular, to the Eddington limit),
 then the cores corresponding to the bifurcating solutions have similar masses. 
The necessary and sufficient conditions for having two cores with significantly 
differing masses, $M_1\ll M_2$, is that the luminosity is much smaller than $L_E$
or the critical luminosity $a$, respectively.

\section{Final remarks.}

In conclusion, we have proven that there exist two radiating systems having 
the same mass, luminosity, asymptotic temperature and surface potential. 
In extreme cases --- for instance when the total mass
is very large and the luminosity low ---  one of the solutions corresponds to a
compact body  having a mass close to the total mass and a small amount of gas,
while the other solution consists of a light compact body enclosed by
a heavy cloud of gas. Returning to the question touched at the beginning  -
one would not make a distinction between a gravastar and a neutron star within the
simple model discussed here.

There is a number of obvious questions that can be asked in connection with
presented results. 

The first question is whether one can justify the assumption  
--- which we make --- of stationary accretion  in the case of a system where
 the cloud of accreting gas is heavier than the compact core. The resolution of this 
problem would require the investigation of the dynamical version of the Shakura model.
The stationary solution of equations (\ref{1} --- \ref{3}) should be inserted 
as initial data to the dynamical equations and if the dynamical solutions remains
close to the stationary data then the assumption would be said to be justified.
We performed an analogous verification of the assumption of stationarity in
the case of selfgravitating gas (in the framework of general relativity) 
(\cite{Kinasiewicz}, \cite{Mach}) with the positive conclusion. The evolving system
remained essential unchanged in its interior for times much smaller than 
 ($ R_\infty / a_\infty$).  We believe that the same conclusion will be valid
for the Shakura model.

The second question is --- accepting that the approximation of steady accretion
is correct --- whether one can reduce the integro-differential problem
to an algebraic one. Our paper provides a proof that this is so if a solution satisfies 
particular boundary data. Numerical examples suggest that the set of solutions that 
satisfy the required boundary conditions is not empty. On the other hand, we believe that
this nonuniqueness of solutions will manifest in a much larger set of steadily
accreting systems which do not necessarily  satisfy our boundary conditions;
that it can be generic. The restriction to a particular sample of systems was made only
in order to simplify the analysis.

Finally, there arises the question that inspired the paper: whether one
can unambiguously distinguish between accretion onto compact bodies from accretion
onto a black hole.  This would require the extension of the above into  
general-relativistic systems. That is, from the general-relativistic  point view, a 
straightforward exercise (see for instance \cite{malec}). There is, however,
a fundamental difficulty (\cite{Thorne}  --- \cite{Rezzola})  in describing 
interaction between accreting perfect
fluid and the   nonisotropic (and thus nonperfect)  outflowing radiation. This 
introduces another level of ambiguity  into the problem, but we believe 
that the  nonuniqueness, that  was discovered  in the newtonian accretion will remain.

This paper has been partially supported by the MNII grant 1PO3B 01229.

\end{document}